\documentclass[english,a4]{elsart}
\usepackage[T1]{fontenc}
\usepackage[latin1]{inputenc}
\usepackage{babel}
\usepackage{graphics}
\usepackage{epsfig}
\usepackage{latexsym}
\usepackage{amsfonts}
\usepackage{array}
\begin{document}
\begin{frontmatter}
\title{Pt-based metallization of PMOS devices for the fabrication of monolithic semiconducting/YBa$ _{2} $Cu$ _{3} $O$_ {7-\delta}$ superconducting devices on silicon}

\maketitle

\author{G.Huot\thanksref{email}}, 
\author{L. Méchin} and
\author{D. Bloyet}

\address{GREYC (CNRS UMR 6072), ENSICAEN \& Université de Caen, 6 Bd du Maréchal Juin, 14050 cedex Caen, France}
\thanks[email]{Corresponding author. email: g.huot@greyc.ismra.fr}
\begin{abstract}
Mo, Pt, Pt/Mo and Pt/Ti thin films have been deposited onto Si and SiO\( _{2} \) substrates by RF sputtering and annealed in the YBa\( _{2} \)Cu\( _{3} \)O\( _{7-\delta } \) growth conditions. The effect of annealing on the sheet resitance of unpatterned layers was measured. A Pt-based multilayered metallization for the PMOS devices was proposed and tested for the monolithic integration of PMOS devices and YBCO sensors on the same silicon substrate. The best results were obtained with a Pt/Ti/Mo-silicide structure showing \( 0.472~\Omega_{\Box} \) interconnect sheet resistivity and $ 2~\times~10^{-4}~\Omega \cdot cm^{2}$ specific contact resistivity after annealing for  \( 60 \)~minutes at~\( 700^{\circ}\)C in \( 0.5 \)~mbar~O\( _{2} \) pressure. 

\end{abstract}
\end{frontmatter}

\section{Introduction}\label{section:1}

The final goal of this work is to combine superconducting YBa\( _{2} \)Cu\( _{3} \)O\( _{7-\delta } \) (YBCO) microbolometers and semiconducting preamplifiers on the same silicon substrate in order to enhance the overall sensor performances. Placing the readout electronics at 77 K as close as possible to the detector has several advantages \cite{Kroger89,Voisi00}. The short distances between sensors and electronics prevent noise pickup from the environment and the cross talk between lines in the case of sensor mutiplexing \cite{Burns93}. Furthermore, the sensor system dimensions are reduced, the electronic white noise can be lowered and the static performances are enhanced \cite{Ghiba92}. Because of the instable nature of YBCO, the superconducting devices have to be processed at the final step. The deposition conditions of high quality YBCO layers on silicon have been described in \cite{Mechin96}. The  semiconducting devices notably have to withstand annealing at temperatures around 700\( ^{\circ }\)C in oxygen atmosphere (during YBCO deposition the substrate holder is held at 700\( ^{\circ }\)C/30~mn/0.5~mBar~O\( _{2} \) pressure followed by a plateau at 500\( ^{\circ } \)C /30~mn/700~mBar O\( _{2} \) pressure). In our experiments a simple academic  PMOS technology was used where we replaced the standard Al metallization buy a multilayer one.
Section \ref{section:2} presents the effects of the annealing temperature and ambient atmosphere on the selected metallic layers (Pt, Mo, Ti) and some combination of them. A Pt-based multilayered metallization system is then proposed and tested in section \ref{section:3}.

\section{ Thermal stability of unpatterned metallic layers \label{section:2}}

We deposited metallic thin films on both  \( 10 \Omega \) Boron-doped silicon and oxidized silicon wafers by RF sputtering. Three pure metal targets were used: Mo, Pt and Ti. The Si and SiO\( _{2} \) substrates were ultrasonically cleaned using organic solutions before loading into the sputtering chamber. The base pressure of the chamber was about \( 10^{-6} \)~mbar. Mo and Pt/Mo layers were deposited onto substrates at room temperature. The Pt and Pt/Ti layers
were prepared in situ onto heated substrates at \( 450^{\circ} \)C
for Ti and \( 550^{\circ} \)C for Pt (see details in \cite{Vilquin02}).
Samples were electrically characterized as-deposited and after annealing for \( 60 \)~minutes either in a \( 10^{-4} \)~mbar vacuum or at \( 0.5 \)~mbar oxygen pressure in the laser ablation chamber used for the YBCO growth.The heater is a radiative resistance and the samples are heated from the back side. The temperature of the substrate holder was fixed at \( 700^{\circ} \)C, which means a sample temperature around \( 750^{\circ} \)C.
The sheet resistance of the metallic layers was measured using the collinear four-probe technique with a HP4156B tester \cite{Schro98}. Results are reported in Tab. \ref{tab:recuit}, where  cross means non-measurable sheet resistance. Qualitative bibliographic data on known interactions with substrate (from \cite{Murar83}) are added for comparison in columns A and B of Tab. \ref{tab:recuit}. Our results are consistent with bibliographic data. We concentrate on the effect of the annealing in oxygen. Only Pt layers show a lower resistance after annealing in O\( _{2} \) both on Si and SiO\( _{2} \) substrates. However, the adhesion of Pt on  SiO\( _{2} \) was not reproducible. The sheet resistance of Pt/Ti layers deposited on SiO\( _{2} \) is not degraded. At last, Pt/Mo on Si could also be a good candidate. From these preliminary measurements it appears that a Pt-based multilayered system should be developed for our PMOS devices in order to withstand the YBCO growth conditions.

\begin{table}
{\raggedright \begin{tabular}{ccccccc}
\hline
\hline 
{\scriptsize Layers}&
{\scriptsize Thickness} &
{\scriptsize As}&
A &
{\scriptsize Interaction} &
B &
{\scriptsize Oxidation}\\
&
&
{\scriptsize deposited}&
&
{\scriptsize with substrates} &
&
{\scriptsize resistance}\\
&
nm&
\( \Omega /\Box  \)&
\( \Omega /\Box  \)&
cf. \cite{Murar83}&
\( \Omega /\Box  \)&
cf. \cite{Murar83}\\
\hline 
Al/SiO\( _{2} \)&
150&
0.113&
\( \times \)&
yes&
\( \times \)&
poor\\
Pt/Si&
270&
17.5&
0.463&
yes&
3.82&
poor\\
{\bf Pt/SiO\( _{2} \)}&
{\bf 270}&
{\bf 16.7}&
{\bf 0.457}&
{\bf no}&
{\bf 0.741}&
{\bf good}\\
Mo/Si&
260&
29.0&
23.2&
yes&
28.8&
good\\
Mo/SiO\( _{2} \)&
260&
22.2&
22.0&
?&
\( \infty  \)&
poor\\
Pt/Ti/Si&
280&
2.92&
\( \times \)&
yes&
\( \times \)&
poor\\
{\bf Pt/Ti/SiO\( _{2} \)}&
{\bf 280}&
{\bf 0.76}&
{\bf 0.72}&
{\bf no}&
{\bf 0.72}&
{\bf good}\\
{\bf Pt/Mo/Si}&
{\bf 310/90}&
{\bf 1.13}&
{\bf 0.550}&
{\bf yes}&
{\bf 0.455}&
{\bf good }\\
Pt/Mo/SiO\( _{2} \)&
260/260&
1.64&
6.25&
no&
\( \times \)&
poor\\
\hline
\hline 

\end{tabular}\par}

\caption{Sheet resistance of layers as-deposited, after annealing at \( 700^{\circ} \)C for \( 60 \)~minutes A: in vacuum (\protect\(10^{-4}\protect \)~mbar) and B: in oxygen atmosphere (\protect\( 0.5\protect \)~mbar O\( _{2} \) pressure). Qualitative data from~\cite{Murar83} on the top layer interactions with the substrates are added in column A and the oxidation resistance property in column B. The possible reliable system good candidates for interconnections are shown in bold. Thickness were measured with an Alpha-Step 200 tool. \label{tab:recuit} }

\end{table}

\section{Complete devices characterization} \label{section:3}

\begin{figure}

{\centering \resizebox{!}{6cm}{\input{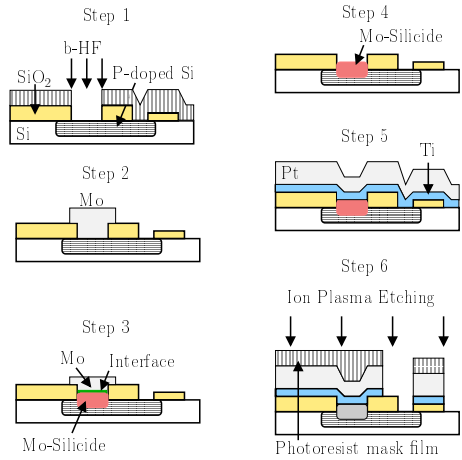}}\par}
{\centering \caption{Schematic sketch of tested metallization process: 1-removal of native oxide on Si contact windows, 2-Sputtering, contact lithography and chemical etching of \( \approx \) 200nm Mo layer, 3-\( 600^{\circ} \)C/\( 60 \) minutes/vacuum annealing, 4-Removal of unreacted Mo, 5-Sputtering of Pt/Ti layers, 6-Lithography of interconnect and gate metallizations and ion etching. \label{fig:process2}}\par}

\end{figure}

Based on above reported preliminary measurements and bibliographic data about near-noble and refractory metal silicides formation (cf. Tab. \ref{tab:Tu81}), we set~up a multilevel metallization process. We decided to form the interconnect and gate metallization with Pt/Ti bilayers (the Ti layer promotes the adhesion of the Pt top layer on SiO\( _{2} \)) and the ohmic-contact onto silicon (drain and source transistor terminals) with Mo-silicide.
The sketch of these operations is illustrated in Fig. \ref{fig:process2}. THe original Al metallization was removed by chemical etching. At step 1, the Si contact windows were cleaned by a HNO\( _{3 }\) buffered HF solution in order to remove the native oxide. In the second step, a 200~nm thick Mo layer was deposited at ambient temperature by sputtering and the contact geometries were defined by photolithography and chemical etching (\( H_{2}SO_{4} \):\( HNO_{3} \):\( H_{2}O \)). An annealing at \( 600^{\circ } \)C for \( 60 \) minutes in vacuum was then performed in order to form the Mo silicide interlayer (step 3). At step 4, unreacted Mo was removed by chemical etching and a Pt/Ti bilayer was sputtered as described in section \ref{section:2}. The contact geometries and the interconnection lines were patterned by photolithography and ion etching (step 6). The YBCO process step
was simulated by annealing the samples at \( 700^{\circ } \)C for \( 60 \)~minutes
in \( 0.5 \)~mbar O\( _{2} \) pressure.

\begin{table}

\begin{tabular}{ccc}

\hline
\hline 
Properties&
Near-noble Metal&
Refractory Metal\\
\hline
1st Phase Formation&
M\( _{2} \)Si&
MSi\( _{2} \)\\
Least Resistive Phase&
MSi&
MSi\( _{2} \)\\
Formation temperature&
\( 200^{\circ}\)C&
\( 600^{\circ}\)C\\
Growth Rate&
\( x^{2}\propto T \)&
\( x\propto T \)\\
Dominant Diffusion Species&
Metal&
Si\\
\hline
\hline 
\end{tabular}

\caption{Comparison of near-noble metal silicides and refractory metal silicides properties from \cite{Tu81}.\label{tab:Tu81}}

\end{table}

\begin{figure}
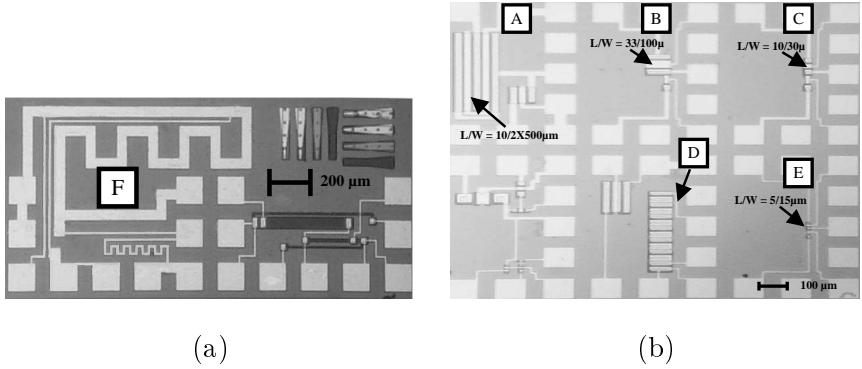

\begin{tabular}{cc}
\resizebox*{5.5cm}{!}{\input{resistance_couche.tex}}&
\resizebox*{5.5cm}{!}{\input{mo122x5.tex}}\\
{\small (a)}&
{\small (b)}\\
\end{tabular}
\caption{Optical photographies of all patterned figures and devices used for samples characterization. In (a), F Patterns are metal lines.  In (b), A, B, C, and E devices are PMOS field-effect transistors, D device is a 14-contacts chain. \label{fig:devices}}
\end{figure}

\begin{figure}
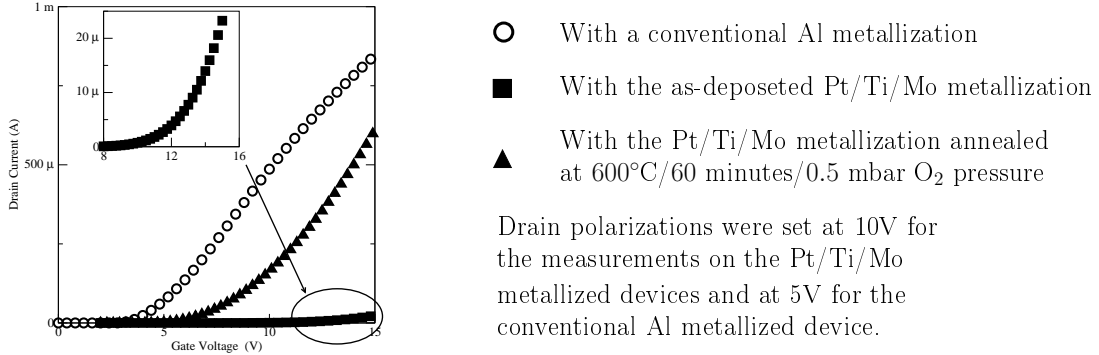

\begin{tabular}{m{6cm}m{6cm}}
\resizebox*{5cm}{!}{\includegraphics{figq1dvgs.eps}}&
\resizebox*{1cm}{!}{\input{legend_figq1dvg.tex}}\\
\end{tabular}
{\centering \caption{ Drain current versus gate voltage curves of the C PMOS transistor device with a Pt/Ti/Mo-silicide metallization system before and after annealing. An Al metallized PMOS Characteristic is added for comparison.  \label{fig:idvg}}\par}
\end{figure}

The metal sheet resistivity was measured on the patterned \( 50~\mu m \) wide and \( 880 \) or \( 3260~\mu m \) long lines (Fig. \ref{fig:devices}a). In fig. \ref{fig:devices}b, one sees the 14-metal/Si contact chain labelled D. The area of each contact
is \( 80~\mu \)m wide (W) and  \(10~\mu  \)m long (L). The contact resistance, \( R_{c} \)
is related to the measured resistance, \( R_{measured} \) by relation~\ref{eq:resistance_specifiquecontact}. \begin{equation}
\label{eq:resistance_specifiquecontact}
R_{c}=\frac{1}{14}\left( R_{measured}-7\times \frac{R_{sheet}\times L}{W}\right) 
\end{equation}
\( R_{sheet} \) is the boron doped silicon resistance per square. The  specific contact resistance, $\rho_{c}=R_{c} \cdot A$, is over-estimated because the geometric areas A are always larger than the effective contact areas. Without the thin Mo silicide layer, we systematically got non-ohmic contacts. Table \ref{tab:contactMoSi2} summarizes the results. Sheet resistance values are consistent with Tab. \ref{tab:recuit} and the specific contact resistivity is one order larger than what is obtained with Al contacts.
We finally present in Fig. \ref{fig:idvg} the \( Id\left(Vgs\right) \) curves of a PMOS transistor (device C on Fig. \ref{fig:devices}b) with a Pt/Ti/Mo-silicide metallization system before and after annealing. The curve for an Al metallized PMOS transistor is reported for comparison. The threshold voltage of our modified device is higher than the Al metallized one. Higher serie resistances and gate metallization resistances  and impurities diffusion into the gate oxyde during our metallization process can explain the high level of the threshold voltage.

\section{Conclusion}

A Pt/Ti/Mo-silicide multilayered system showed \( 0.472 \Omega/\Box \) interconnect sheet resitivity and \( 2~10\times^{-4}~\Omega \cdot \)cm\( ^{2} \) specific contact resistivity after annealing for \( 60 \)~minutes at \( 700^{\circ } \)C in \( 0.5 \)~mbar O\( _{2} \) pressure. PMOS transistors were fabricated with such a metallization procedure. Altough the device characteristics don't be similar to the conventional ones, thus demonstrating a first step of a monolithic integration of semiconducting/YBCO superconducting devices on silicon. The next technological step to be demonstrated is the interconnections between a semiconducting device and a superconducting one.

\section*{Acknowledgments}

We present our acknowledgments for their technical help to G. LeRhun,
L. Pichon and S. Eimer and we thank the GM Rennes Laboratory for supplying the
devices used in this study.

\begin{table}
{\raggedright \begin{tabular}{ccccc}
\hline
\hline 
Steps (cf. fig. \ref{fig:process2})&
2&
3&
6&
After annealing\\
&
&
&
&
 $700^{\circ}$C/$60$~mn/$0.5$~mbar O$_{2}$ pressure\\
\hline 
 R\( _{sheet} \) ($ \Omega/\Box $)&
-&
-&
0.569&
0.472\\
 \(\rho _{c} \) ($ \Omega \cdot cm^{2} $)&
Non-ohmic&
0.01&
\( 4\times 10^{-4} \)&
\( 2\times 10^{-4} \)\\
\hline
\hline 
\end{tabular}\par}
{\centering \caption{Comparison of the sheet resistivity (R\( _{sheet} \)) and the specific contact resistivity (\(\rho _{c} \)) at each step of the metallization process described in Fig. \ref{fig:process2}. \label{tab:contactMoSi2}}\par}
\end{table}

\bibliographystyle{plain}
\bibliography{mam03-5}

\begin{thebibliography}{1}

\bibitem{Burns93}
M.~J. Burns, K.~Char, B.~F. Cole, W.S. Ruby, and S.A. Satchjen.
\newblock {\em Appl. Phys. Lett}, 12:1435--143, 1993.

\bibitem{Ghiba92}
G.~Ghibaudo, F.~Balestra, and A.~Emrani.
\newblock {\em MicroElectron. Eng.}, 19:833--840, 1992.

\bibitem{Kroger89}
H.~Kroger, C.~Hilbert, D.~A. Gibson, U.~Ghoshal, and L.~N. Smith.
\newblock {\em Proc. IEEE}, 77(8):1287--1301, 1989.

\bibitem{Mechin96}
L.~Méchin, J.-C. Villégier, G.~Rolland, and F.~Laugier.
\newblock {\em Physica C}, 269(1):124--130, 1996.

\bibitem{Murar83}
S.P. Murarka.
\newblock {\em Silicides for \textsc{VLSI} Applications}.
\newblock 1983.

\bibitem{Schro98}
D.~K. Schroder.
\newblock {\em Semiconductor Material and Device Characterization}.
\newblock Wiley-InterSci., 2nd edition edition, 1998.

\bibitem{Tu81}
K.~N. Tu.
\newblock {\em J. Vac. Sci. Technol. A}, 19(3), 1981.

\bibitem{Vilquin02}
B.~Vilquin, G.~Le Rhun, R.~Bouregba, G.~Poullain, and H.~Murray.
\newblock {\em App. Surf. Science}, 191(1):63--73, 2002.

\bibitem{Voisi00}
F.~Voisin, G.~Klisnick, Y.~Hu, M.~Redon, J.~Delerue, A.~Gaugue, and
  A.~Kreisler.
\newblock {\em 4$^{th}$ Workshop On Low Temperature Electronice},
  W\textsc{PP}--171:131, 2000.

\end{thebibliography}

\end{document}